\documentstyle[aps,prl,epsf]{revtex}
\draft

\newcommand{\etal}{{\it et al.}}

\def\BE{\begin{equation}}
\def\EE{\end{equation}}
\def\BEA{\begin{eqnarray}}
\def\EEA{\end{eqnarray}}
\def\EL{\nonumber\\}

\def\Dslash{\mathop{\not\!\! D}}

\begin{document}
\twocolumn[\hsize\textwidth\columnwidth\hsize\csname
@twocolumnfalse\endcsname

\title{ Scaling tests of the improved Kogut-Susskind quark action}


\author{
Claude~Bernard$\null^a$,
Tom Burch$\null^b$,
Thomas~A.~DeGrand$\null^c$,
Carleton~DeTar$\null^d$,
Steven~Gottlieb$\null^e$,
Urs~M.~Heller$\null^f$,
James~E.~Hetrick$\null^g$,
Kostas Orginos$\null^b$,
Bob~Sugar$\null^h$
and 
Doug~Toussaint$\null^b$
}
\address{
$\null^a$Department of Physics, Washington University, St.~Louis, MO 63130, USA,
$\null^b$Department of Physics, University of Arizona, Tucson, AZ 85721, USA,
$\null^c$Physics Department, University of Colorado, Boulder, CO 80309, USA
$\null^d$Physics Department, University of Utah, Salt Lake City, UT 84112, USA,
$\null^e$Department of Physics, Indiana University, Bloomington, IN 47405, USA,
$\null^f$SCRI, Florida State University, Tallahassee, FL 32306-4130, USA,
$\null^g$Department of Physics, University of the Pacific, Stockton, CA 95211-0197, USA,
$\null^h$Department of Physics, University of California, Santa Barbara, CA 93106, USA,
}
\date{\today}

\maketitle

\begin{abstract}\noindent
Improved lattice actions for Kogut-Susskind quarks have been shown to
improve rotational symmetry and flavor symmetry.  In this work we
find improved scaling behavior of the $\rho$ and nucleon masses expressed
in units of a length scale obtained from the static quark potential, and better
behavior of the Dirac operator in instanton backgrounds.
\end{abstract}
]

%
\narrowtext

Because of their large computational requirements,
full QCD lattice simulations profit greatly from the use of
improved actions, which allow better physics to be extracted
from simulations at moderate lattice spacings.
The Kogut-Susskind formulation of lattice fermions is attractive
for full QCD simulations because the thinning of the degrees of
freedom reduces the computational effort and, more importantly, because
the residual unbroken chiral symmetry prevents additive renormalization
of the quark mass, and hence eliminates problems with exceptional
configurations.

The lattice artifacts in the Kogut-Susskind formulation are order
$a^2$, unlike the Wilson formulation which has an order $a$ artifact,
which can be cancelled by the ``clover'' improvement.
A third nearest neighbor coupling introduced by Naik cancels
order $a^2$ violations of rotational symmetry in the free quark
propagator\cite{NAIK}, and this has been shown to lead to an improvement
in the rotational symmetry of meson propagators\cite{MILC_NAIK}.
Another lattice artifact is the breaking of flavor symmetry, signalled
by the fact that only one of the pions produced from the four flavors
of quarks has an exactly vanishing mass at zero quark mass.
This flavor symmetry breaking
can be understood as a scattering of a quark from one corner
of the Brilloin zone to another through the exchange of a gluon
with momentum near $\pi/a$\cite{LEPAGE_TSUKUBA}.  Roughly speaking,
the cure for this problem consists of introducing a form factor for
the quark-gluon interaction by smearing out, or ``fattening'' the
gauge connection in the quark action.  In our previous works we
have investigated the effects of different fat link actions on 
the flavor symmetry violation seen in the pion mass
spectrum\cite{MILC_FAT,MILC_FATTER,MILC_FATTEST}.  (see also
Ref.~\cite{ILLINOIS_FAT}.)
A theoretically attractive action is the ``Asqtad'' action studied
in Ref~\cite{MILC_FATTEST}, which includes the Naik correction to improve 
rotational symmetry, fattening of the nearest neighbor coupling to
cancel couplings to gluons with any momentum component equal to
$\pi/a$, and a term introduced by Lepage\cite{LEPAGE98}
which corrects order $a^2$
errors at low momentum introduced by the form factor.
This ``Asqtad'' action cancels all tree level order $a^2$
errors\cite{LEPAGE98}.

Our previous works tested improvements in physical quantities 
that were directly related
to the improvements in the action.  That is, the Naik term is
introduced to improve rotational symmetry in the quark propagator and
was seen to improve rotational symmetry in the meson propagators.  The
fat link term was designed to reduce flavor symmetry breaking and was
seen to reduce the flavor symmetry breaking in the pion mass spectrum.
It is perhaps less obvious that other quantities will be improved, and
such tests are the subject of this note.  We have calculated hadron
masses and the static quark potential at different lattice spacings, and
we can compare the scaling of these quantities with results with other
actions.  We have also investigated the physics of instantons on the
lattice by computing eigenvalues of the improved and conventional
Kogut-Susskind Dirac operators on smooth instanton configurations and
on semi-realistic ``noisy'' configurations containing an instanton.


We test scaling by computing hadron masses using a length scale
determined from the static quark potential.  In particular, we will use
a variant of the Sommer parameter defined by $r_1^2 F(r_1) = 1.00$,
where the commonly used $r_0$ is defined by $r_0^2 F(r_0)=1.65$.  (The
advantage of $r_1$ is that it is determined at a shorter length scale,
where the potential can be determined more accurately, and so
lattice spacings in simulations with different parameters can be matched
more accurately using $r_1$.)  For the quenched potential, this is
related to $r_0$ and $\sigma$ by $r_1/r_0 = 0.725$ and 
$r_1 \sqrt{\sigma} = 0.85$.  For quenched simulations with the single plaquette
gauge action, we take the lattice spacing from the interpolating
formula of Guagnelli, Sommer and Wittig\cite{GUAGNELLI}:
\BEA
  \log(a/r_1) &=& -1.3589 - 1.7139 (\beta-6) \EL
     &+& 0.8155 (\beta-6)^2 - 0.6667 (\beta-6)^3 \ \ ,
\EEA
while for quenched simulations with the Symanzik improved gauge action
we fit a similar formula to the string tension results of
Collins \etal\cite{SCRIWILSON}
and our own results at $\beta=10/g^2=8.0$
\BEA
  \log(a/r_1) &=& -0.970 - 0.840 (\beta-8) \EL
      &+& 0.413 (\beta-8)^2 + 0.304 (\beta-8)^3 \ \ \ ,
\EEA
with an error of around 1\%.
We use the Goldstone pion mass to adjust the quark mass in the various
simulations, interpolating to the point where $m_\pi r_1 = 0.778$.
(This somewhat arbitrary value corresponds to $a m_q=0.02$ at
$10/g^2=8.0$.)

\begin{figure}[tb]
\epsfxsize=3.5in
\epsfysize=3.5in
\epsfbox[0 0 4096 4096]{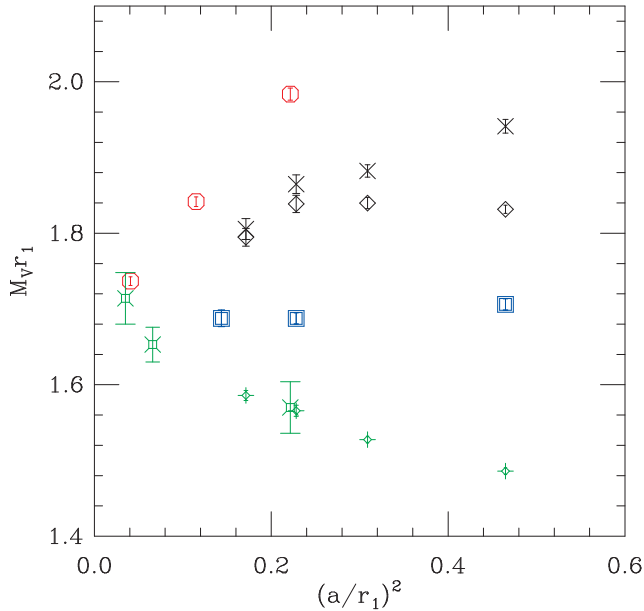}
\caption{
\label{RHOFIG}
Rho masses in units of $r_1$.  The meaning of the symbols is described
in the text.
}
\end{figure}

\begin{figure}[tb]
\epsfxsize=3.5in
\epsfysize=3.5in
\epsfbox[0 0 4096 4096]{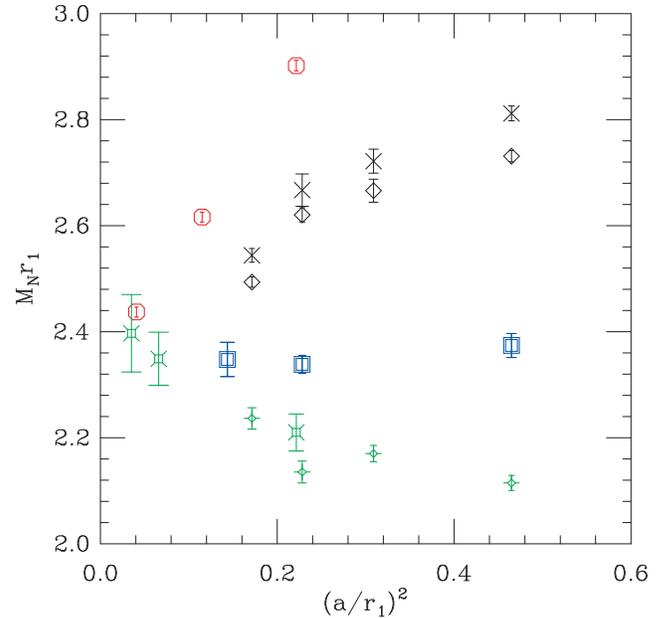}
\caption{
\label{NUCFIG}
Nucleon masses in units of $r_1$.  The symbols are the same as in
Fig.~\protect\ref{RHOFIG}.
}
\end{figure}

We calculated the quenched hadron spectrum using the improved quark
action and Symanzik improved gauge action at $10/g^2=7.4$ and $7.75$ using
$16^3\times 32$ lattices archived in our earlier work\cite{MILCSYM},
and on $20^3\time 64$ lattices at $10/g^2=8.0$ generated in our current project,
which have a lattice spaing of $a \approx 0.14$ fm. 
In Fig.~\ref{RHOFIG} we plot the vector meson mass in units
of $r_1$ versus the squared lattice spacing for several combinations
of gauge and quark actions.
The
bold	
squares are from our improved quark action, with the Symanzik improved
gauge action.
The octagons are from a simulation with the conventional staggered
action, and the simple one-plaquette gauge action\cite{MILCQUENCHED}.
The diamonds are the conventional staggered quark action, with a
Symanzik improved gauge action, and the crosses are the Symanzik
improved gauge action with a quark action containing only the Naik
improvement\cite{MILC_NAIK,MILCSYM}.
We also show some results for tadpole improved clover Wilson quark actions.
The plusses are from the SCRI collaboration\cite{SCRIWILSON},
using Symanzik improved gauge action, and the fancy squares from
the UKQCD collaboration\cite{UKQCDWILSON}, using the one-plaquette gauge
action.
Fig.~\ref{NUCFIG} is a similar plot of the nucleon masses in units of
$r_1$.
In both of these plots the ``Asqtad'' action shows scaling behavior
that is dramatically better than the other actions tested.

Instantons play an important role in the Euclidean description of
QCD. They provide the solution to the $U(1)$
problem~\cite{tHooft2}, explain chiral symmetry
breaking~\cite{Shuryak}, and at least at a qualitative level reproduce
the low mass hadron correlators~\cite{Grandy,Negele99}. Thus it is
important that the lattice actions we use to simulate QCD approximate
the continuum behavior at finite lattice spacing. Here we
study the topological aspects of the improved Kogut-Susskind actions
and compare with the standard formulations.  The effect of topology on
the spectrum of the Dirac matrix has been also studied
in~\cite{Heller99}, and in~\cite{KogutTopo98}. In~\cite{Heller99}, the
microscopic spectral density was computed and shown that Kogut-Susskind
fermions are insensitive to topology on coarse lattice spacings
($a \sim 0.5$ fm). In ref.~\cite{KogutTopo98} it was shown that at 
lattice spacing about 0.07 fm there is a clear separation between topological and
non-topological modes.

\begin{figure}[tbh]
\epsfxsize=3.5in
\epsfysize=3.5in
\epsfbox[0 0 4096 4096]{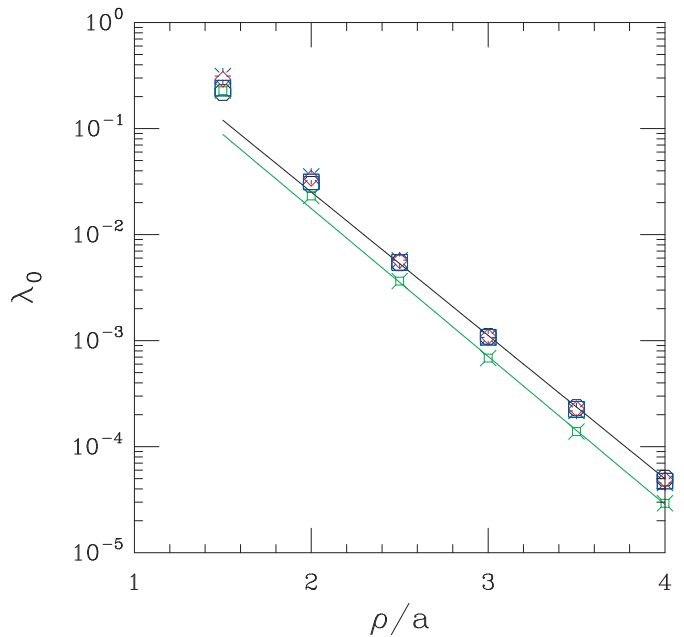}
\caption{
\label{INST_SIZE_FIG}
The small eigenvalue of the Dirac operator as a function
of instanton size for several actions.  The octagons are the one link action,
the squares are the Naik action, the diamonds are the ``Fat7'' action,
the bursts are the `Fat7 Naik'', and the fancy squares are the Asqtad action.
}
\end{figure}

We first studied the behavior of the low eigenvalues of the Dirac
matrix in the background of an instanton
by computing eigenvalues
and eigenvectors of $-\Dslash_e^2$ using the Ritz functional
technique~\cite{KALKREUTER99}.  $\Dslash_e^2$ is the squared Dirac
operator restricted to even lattice sites.  In an instanton background
$\Dslash_e^2$ should have two chiral eigenvectors with small eigenvalues,
each corresponding to two eigenvectors of $\Dslash$.
We computed the eight lowest eigenvalues of 
$-\Dslash^2$ on smooth instantons. The instantons are put on the
lattice by discretizing the continuum formula for the instanton gauge
field $A_\mu$\cite{DKH98}, and the resulting gauge configuration
was smoothed by twenty APE smearing sweeps.
We calculated eigenvectors on lattices ranging in size from $4^4$ to $16^4$
containing an instanton with radius $\rho$ equal
to one fourth the lattice size, $\rho = L/4$.
Since there is
no QCD dynamics to define a length scale on these smoothed lattices,
changing $\rho/a$ can equally well be
considered to be varying the size of the instanton at fixed lattice
spacing or varying the lattice spacing for a fixed instanton size.
We computed the eight lowest eigenvalues of $-\Dslash^2$ for the
standard KS action, the Naik action, the ``Fat7'' action( all couplings to
gluons with momentum $\pi/a$ set to zero), the ``Fat7'' with the Naik
term, and finally the Asqtad action (with $u_0=1$ since tadpole improvement
has no meaning on smooth backgrounds)\cite{MILC_FATTEST}.
As expected we found two small
eigenvalues, which approach zero as the lattice spacing goes
to zero.
These two eigenvalues were degenerate to the accuracy of the
computation.  In Fig.~\ref{INST_SIZE_FIG} we plot the small eigenvalue
$\lambda$ of $\Dslash$ for all the actions tested as a function of
$\rho/a$.  For all the actions, the small eigenvalues go
to zero exponentially as $a/\rho$ decreases.
Neither the Naik term nor the fattening of the link term affects the
eigenvalue significantly.  (This is not surprising --- there is no point in
smoothing out the link when you are already working on a smooth background.)
However, the Asqtad action, which differs from the Fat7 action in that
it contains the small momentum correction introduced by Lepage, has 
smaller eigenvalues than the other actions.
We also
measured the chiralities $\chi=\psi^\dagger\gamma_5\psi$ of the
eigenvectors associated with the small eigenvalues. Although $\chi$
differs significantly from $-1$, it differs from $-1$
by an amount proportional to $a^2/\rho^2$.

\begin{figure}[tb]
\epsfxsize=3.5in
\epsfysize=3.5in
\epsfbox[0 0 4096 4096]{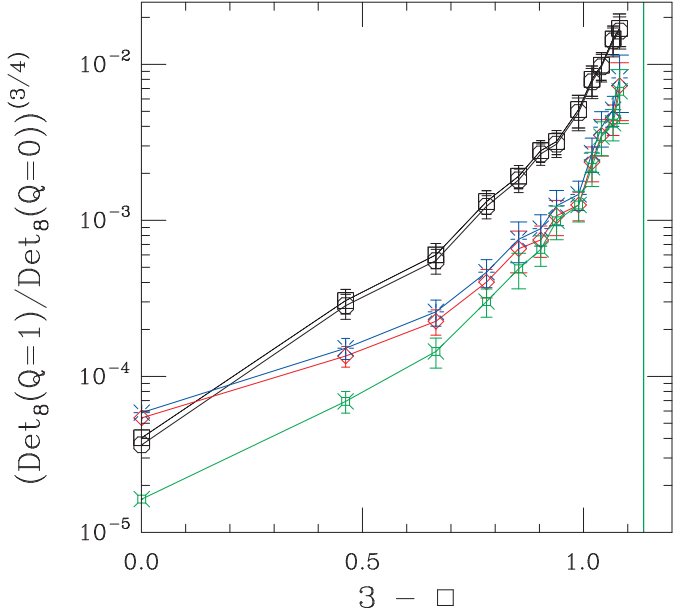}
\caption{
\label{INST_HEAT_FIG}
The approximate instanton suppression factor for three quark flavors as a function
of the amount of ``heating'' applied to the instanton.  The meaning of the
symbols is the same as in Fig.~\protect\ref{INST_SIZE_FIG}
}
\end{figure}

To see if these small eigenvalues persist when the QCD dynamics
is turned on, we ``heated''
the $\rho/a=2$ instanton on the  $8^4$ lattice.
This was done by 
short quenched molecular dynamics trajectories, using the 
tadpole improved one loop Symanzik
gauge action with $\beta=8.0$.
Since we want to introduce short distance structure without disturbing
the long distance topological structure of the initial lattice,
we used short trajectories --- ten molecular dynamics steps of size $0.02$.
We ran for $20$ such
trajectories, saving the lattice at the end of each
trajectory.
The resulting sequence of $20$ lattices
was discarded if any of the lattices had topological charge
different frome one. In total we produced 31 different heating runs
in which the instanton survived. At the end of the $20$th
trajectory the average plaquette was $1.92$, similar to the
average $1.86$ of thermalized $\beta=8.0$ quenched lattices.
For comparison, we did 24 runs starting from smooth lattices containing
no instantons.
In Fig.~\ref{INST_HEAT_FIG} we plot the 3/4 power
of the averaged product of the eight smallest
eigenvalues of $\Dslash_e^2$ for the $Q=1$ heating runs, divided by the same
quantity from the $Q=0$ runs.
For all of the quark actions, $u_0$ was kept at one during these
heating runs.
This quantity, ``${\rm Det}_8^{3/4}$'', is an approximation to the
factor by which three flavors of massless dynamical quarks would
suppress such instanton configurations
in a full QCD simulation.
The abscissa
in Fig.~\ref{INST_HEAT_FIG} is 
$3 - \langle\sum_{\mu\nu}{\rm Tr}P_{\mu\nu}\rangle$,
where $P_{\mu\nu}$ is the plaquette.
This is related to the amount of disorder induced by the heating
The vertical line marks the average plaquette of a 
thermalized lattice.
As the amount of disorder increases, the size
of the small eigenvalues in the $Q=1$ runs
increases, but the eigenvalues of the fat
link $\Dslash$ rise less than those of the Naik and
standard KS actions.
Roughly speaking, Fig.~\ref{INST_HEAT_FIG}
shows that at $a \approx 0.14$ fm
three dynamical quarks with fat link actions suppress
instantons by more than a factor of two relative to the conventional action.
We have also looked at the chirality of the
would-be zero modes.  Again, as the disorder increases the deviation
from $-1$ also increases, but the improved action holds up better.
After 20 heating trajectories with the $\rho=2a$ instanton, the
chirality for the fat actions is about $-0.4$ while for the standard
KS action (and the Naik) it is $-0.2$.
Finally, the degeneracy of the two
would-be zero modes of $-\Dslash^2$ is lifted as disorder is
added. The splitting of these modes is related to the
flavor symmetry breaking and, as expected from our previous
studies~\cite{MILC_FATTEST}, is about a factor of three smaller for the
fat actions than for the standard Kogut-Susskind action.

Lastly, we looked at the eight low eigenvalues of $-\Dslash_e^2$
for the Asqtad and standard Kogut-Susskind actions on 424 thermalized
$8^4$ quenched $\beta=8.0$ lattices,
which have a lattice spacing of about 0.14 fm.
At this volume we have $1$ to $2$
topological objects of average size $\rho = 2a$ to $3a$ per lattice,
based on the instanton distribution measurements in~\cite{DKH98}.
Typically, when the topological
charge is non-zero we find small eigenvalues with high
absolute value of the chirality ($0.2$ to $0.3$) for the Asqtad
action. In a chirality versus eigenvalue scatter plot the topological
eigenvalues clearly separate from the non-topological. Similar but 
weaker separation is also observed for the standard
Kogut-Susskind action.  We have also 
checked the validity of the index theorem. For the Kogut-Susskind
fermions which represent $4$ flavors it takes the form
$4\,Q = N_- - N_+$,
where $N_-$ are the number of left-handed modes, $N_+$ the number of
right-handed modes and $Q$ is the topological charge.
On the lattice this relation does not hold exactly for staggered fermions,
but the quantities $4\,Q$ and $N_- - N_+$ are strongly correlated.
For the Asqtad action this correlation is $90\%$, while for the standard
Kogut-Susskind is $83\%$. In both cases the topological charge was measured
after $200$ APE smearings.


The code used for generating smooth instanton configurations and for
measuring the topological charge was written by Anna Hasenfratz and Tamas
Kovacs.  
Computations were done on the T3E and PC cluster at NERSC, on the T3E and SP2
at SDSC, on the NT cluster and Origin 2000 at NCSA, and on the Origin
2000 at BU.
This work was supported by the U.S. Department of Energy under contracts
DOE -- DE-FG02-91ER-40628,	
DOE -- DE-FG03-95ER-40894,	
DOE -- DE-FG02-91ER-40661,	
DOE -- DE-FG05-96ER-40979 	
and
DOE -- DE-FG03-95ER-40906 	
and National Science Foundation grants
NSF -- PHY99-70701 		
and
NSF -- PHY97--22022.    	

\end{document}